\documentstyle[epsfig,psfig]{texas}

\def \be{\begin{equation}}
\def \ee{\end{equation}}
\def \bea{\begin{eqnarray}}
\def \eea{\end{eqnarray}}

\def \s2{\sqrt 2}

\begin{document}

\title{Non-linear Evolution of Rotating Relativistic Stars}

\author{Nikolaos STERGIOULAS$^{1}$, Jos{\'e} A. FONT$^{1}$ and Kostas D. KOKKOTAS$^{2}$}
 
\address{(1) Max-Planck-Institut f{\"u}r Gravitationsphysik \\
Albert-Einstein-Institut \\
Schlaatzweg 1, D-14473, Potsdam, Germany \\
}

\address{(2) Department of Physics \\ 
Aristotle University of Thessaloniki \\
Thessaloniki 54006, Greece \\
{\rm Email: niksterg@aei-potsdam.mpg.de, font@aei-potsdam.mpg.de, \\ 
kokkotas@astro.auth.gr}
}

\begin{abstract}
  We present first results of the non-linear evolution of rotating
  relativistic stars obtained with an axisymmetric relativistic
  hydrodynamics code in a fixed spacetime.  As initial data we use
  stationary axisymmetric and perturbed configurations. We find that,
  in order to prevent (numerical) angular momentum loss at the surface
  layers of the star a high-resolution grid (or a numerical scheme
  that retains high order at local extrema) is needed. For
  non-rotating stars, we compute frequencies of radial and non-radial
  small-amplitude oscillations, which are in excellent agreement with
  linear normal mode frequencies computed in the Cowling
  approximation. As a first application of our code, quasi-radial
  modes of rapidly rotating relativistic stars are computed.  By
  generalizing our numerical code to 3-D, we plan to study the
  evolution and non-linear dynamics of toroidal oscillations
  ($r$-modes) of rapidly rotating neutron stars, which are a promising
  source of gravitational waves.
\end{abstract}

\section{Introduction}
\label{intro}

The numerical evolution of neutron stars in full General Relativity
has been the focus of many research groups in recent years
\cite{font2,shibata,mathews,nakamura}. So far, these studies have
been limited to initially non-rotating stars.  However, the numerical
investigation of many interesting astrophysical applications, such as
the rotational evolution of proto-neutron stars and merged neutron
stars or the simulation of gravitational radiation from unstable
pulsation modes, requires the ability of accurate long-term evolutions
of rapidly rotating stars. We thus present here the first study of
hydrodynamical evolutions of rotating neutron stars in the
approximation of a static spacetime.  This approximation allows us to
evolve relativistic matter for a much longer time than present coupled
spacetime plus hydrodynamical evolution codes. Since the pulsations of
neutron stars are mainly a hydrodynamical process, the exclusion of
the spacetime dynamics has only a limited effect and allows for
qualitative conclusions to be drawn.

The rotational evolution of neutron stars can be affected by several
instabilities (see \cite{nik1} for a recent review). If hot
protoneutron stars are rapidly rotating, they can undergo a dynamical
bar-mode instability \cite{centrella}. When the neutron star has
cooled to about $10^{10}$K after its formation, it can be subject to
the Chandrasekhar-Friedman-Schutz instability \cite{CFS1,CFS2} and it
becomes an important source of gravitational waves. It was recently
found that the $l=m$ $r$-mode has the shortest growth time of the
instability \cite{Andersson1,Friedman} and it can transform a rapidly
rotating newly-born neutron star to a Crab-like slowly-rotating pulsar
within about a year after its formation \cite{Andersson2,Lindblom}. In
this model, there are two important questions still to be answered
\cite{Owen}: What is the maximum amplitude that an unstable $r$-mode
can reach (limited by nonlinear saturation) and is there any transfer
of energy to other stable or unstable modes via non-linear couplings?
Such questions cannot be answered by computations of normal modes of
the linearized pulsation equations, but require non-linear effects to
be taken into account. We therefore need to develop the capability of
full non-linear numerical evolutions of rotating stars in General
Relativity.

Our present 2-D (axisymmetric) code uses high-resolution
shock-capturing (HRSC) finite-difference schemes for the numerical
integration of the general relativistic hydrodynamic equations
\cite{Leveque92} (see \cite{jcam} for a recent review of applications
of HRSC schemes in relativistic hydrodynamics). In a similar context
to the one presented here let us note that such schemes have been
succesfully used before in the study of the numerical evolution and
gravitational collapse of non-rotating neutron stars in 1-D
\cite{Romero}.  An alternative approach, based on pseudospectral
methods, has been presented in \cite{eric1}.  

Using our code in 1-D time-evolutions we can accurately identify
specific normal modes of pulsation. In 2-D the code is suitable for
the evolution of rotating stars, with the additional complication of
having to pay special attention to an angular momentum-loss at the
(non-spherical) surface of the star, as we will show below.

\section{Initial Configurations}
\label{initial}

Our initial models are fully relativistic, stationary and axisymmetric
configurations, rotating with uniform angular velocity $\Omega$. The
metric in quasi-isotropic coordinates \cite{Butterworth} is
\begin{eqnarray}
ds^2=-e^{2\nu}dt^2 + B^2e^{-2\nu}r^2\sin^2\theta(d\phi-\omega dt)^2
+e^{2\alpha}(dr^2+r^2d\theta^2),
\label{metric}
\end{eqnarray}
where $\nu$, $B$, $\alpha$ and $\omega$ are metric functions 
(gravitational units
are implied). In the non-rotating limit the above metric reduces to
the metric of spherical relativistic stars in isotropic coordinates.

We assume a perfect fluid, zero-temperature equation of
state (EOS), for which the energy density is a function of pressure
only. The following relativistic generalization of the Newtonian 
polytropic EOS is chosen:
\begin{eqnarray}
p&=&K\rho_0^{1+1/N}
\\
\epsilon&=& \rho_0 +Np,
\end{eqnarray}
\noindent
where $p$ is the pressure, $\epsilon$ is the energy density, $\rho_0$ is the
rest-mass density, $K$ is the polytropic constant and $N$ is the
polytropic exponent.

The initial equilibrium models are computed using a numerical code by
Stergioulas \& Friedman \cite{nik2} which follows the Komatsu,
Eriguchi \& Hatchisu \cite{keh} method (as modified in
\cite{cst}) with some changes for improved accuracy (see \cite{NSGE}
for a comparison with other codes). The code is freely available and
can be downloaded from the following URL address:
http://www.gravity.phys.uwm.edu/Code/rns.

\section{Relativistic Hydrodynamic Equations}
\label{hydro}

The equations of (ideal) relativistic hydrodynamics are obtained from
the local conservation laws of density current, $J^{\mu}$ and
stress-energy, $T^{\mu\nu}$
\begin{eqnarray}
\nabla_{\mu}J^{\mu}&=&0
\\
\nabla_{\mu}T^{\mu \nu}&=&0
\end{eqnarray}
\noindent
with
\begin{eqnarray}
J^{\mu}&=&\rho_0u^{\mu}
\\
T^{\mu \nu}&=&\rho_0hu^{\mu}u^{\nu} + p g^{\mu \nu},
\end{eqnarray}
\noindent
for a general EOS $p=p(\rho,\varepsilon)$. This choice of the
stress-energy tensor limits our study to perfect fluids.

In the previous expressions $\nabla_{\mu}$ is the covariant derivative,
$u^{\mu}$ is the fluid 4-velocity and $h$ is the specific enthalpy
\begin{eqnarray}
h=1+\varepsilon+\frac{p}{\rho_0}
\end{eqnarray}
\noindent
with $\varepsilon$ being the specific internal energy, related to
the energy density $\epsilon$ by
\begin{eqnarray}
\varepsilon=\frac{\epsilon}{\rho_0}-1.
\end{eqnarray}
\noindent

With an appropriate choice of matter fields the equations of
relativistic hydrodynamics constitute a (non-strictly) hyperbolic
system and can be written in a flux conservative form, as was first
shown in \cite{mim} for the
one-dimensional case.  The knowledge of the characteristic fields of
the system allows the numerical integration to be performed by means
of advanced high-resolution shock-capturing (HRSC) schemes,using
approximate Riemann solvers (Godunov-type methods).  The
multidimensional case was studied in \cite{betal97},
within the framework of the 3+1 formulation.  Further
extensions of this work to account for {\it dynamical} spacetimes,
described by the full set of Einstein's non-vacuum equations, can be
found in \cite{font2}. Fully {\it covariant}
formulations of the hydrodynamic equations (i.e., not restricted to
{\it spacelike} approaches) and also adapted to Godunov-type methods,
are presented in \cite{pf,fp}.

In the present work we use the hydrodynamic equations as formulated in
\cite{betal97}. Specializing for the metric given by
Eq.~(\ref{metric}), the 3+1 quantities read
\begin{eqnarray}
\tilde{\alpha} &=& e^{\nu}
\\
\beta_{\phi} &=& -\omega B^2 e^{-2\nu} r^2 \sin^2\theta
\\
\gamma_{rr} &=& e^{2\alpha}
\\
\gamma_{\theta\theta} &=& r^2 e^{2\alpha}
\\
\gamma_{\phi\phi} &=& B^2 e^{-2\nu} r^2 \sin^2\theta
\end{eqnarray}
\noindent
where $\tilde{\alpha}$ is the lapse function (the tilde is
used to avoid confussion with the metric potential $\alpha$) and
$\beta_{\phi}$ is the azimuthal shift.

The hydrodynamic equations are written as a first-order flux
conservative system of the form
\begin{eqnarray}
\frac{\partial {\bf u}}{\partial t} +
\frac{\partial \tilde{\alpha} {\bf f}^r} {\partial r} +
\frac{\partial \tilde{\alpha} {\bf f}^{\theta}} {\partial \theta} =
{\bf s}
\label{hydro_system}
\end{eqnarray}
\noindent
where ${\bf u}, {\bf f}^r, {\bf f}^{\theta}$ and ${\bf s}$ are,
respectively, the state vector of evolved quantities, the radial and polar
fluxes and the source terms. More precisely, they take the form
\begin{eqnarray}
{\bf u} &=& (D,S_r,S_{\theta},S_{\phi},\tau)
\\
{\bf f}^r &=& (Dv^r, S_rv^r+p, S_{\theta}v^r, S_{\phi}v^r, (\tau+p)v^r)
\\
{\bf f}^{\theta} &=& (Dv^{\theta}, S_rv^{\theta}, S_{\theta}v^{\theta}+p,
S_{\phi}v^{\theta}, (\tau+p)v^{\theta}).
\end{eqnarray}
\noindent
The source terms can be decomposed in the following way
\begin{eqnarray}
{\bf s} = \tilde{\alpha} {\bf s}^{\star} - 
\tilde{\alpha} {\bf f}^r \frac{\partial \log\sqrt{\gamma}}{\partial r}
-\tilde{\alpha} {\bf f}^{\theta} \frac{\partial \log\sqrt{\gamma}}
{\partial \theta}
\end{eqnarray}
\noindent
with $\gamma=\det\gamma_{ij}$ and
\begin{eqnarray}
{\bf s}^{\star} = \left(0, T^{\mu\nu}\left[\frac{\partial g_{\nu j}}
{\partial x^{\mu}} - \Gamma^{\delta}_{\mu\nu}g_{\delta j}\right],
\tilde{\alpha}\left[T^{\mu t}\frac{\partial\log\tilde{\alpha}}
{\partial x^{\mu}} - T^{\mu\nu}\Gamma^t_{\mu\nu}\right]\right)
\end{eqnarray}
\noindent 
with $j=r,\theta,\phi$. The definitions of the evolved quantities in terms
of the ``primitive" variables ($\rho,v_j,\varepsilon$) are
\begin{eqnarray}
D &=& \rho_0 W
\\
S_j &=& \rho_0 h W^2 v_j
\\
\tau &=& \rho_0 h W^2 - p - D
\end{eqnarray}
\noindent
where $W$ is the relativistic Lorentz factor
\begin{eqnarray}
W \equiv \tilde{\alpha} u^t = \frac{1}{\sqrt{1-v^2}}
\end{eqnarray}
\noindent
with $v^2=\gamma_{ij}v^iv^j$. The 3-velocity components are obtained
from the spatial components of the 4-velocity in the following way
\begin{eqnarray}
v^i=\frac{u^i}{W}+\frac{\beta^i}{\tilde{\alpha}}.
\end{eqnarray}
Explicit expressions for the
non-vanishing Christoffel symbols for metric~(\ref{metric}), appearing
in the source terms of the hydrodynamic equations, are presented in \cite{fsk}.

\section{Numerical Methods}

As stated before, our numerical integration of system
(\ref{hydro_system}) is based on Godunov-type methods (also known as
HRSC schemes). In a HRSC scheme, the knowledge of the characteristic
fields (eigenvalues) of the equations, together with the corresponding
eigenvectors, allows for accurate integrations, by means of either
exact or approximate Riemann solvers, along the fluid characteristics.
These solvers, which constitute the kernel of our numerical algorithm,
compute, at every interface of the numerical grid, the solution of
local Riemann problems (i.e., the simplest initial value problem with
discontinuous initial data). Hence, HRSC schemes automatically
guarantee that physical discontinuities appearing in the solution,
e.g., shock waves, are treated consistently (the {\it shock-capturing}
property).  HRSC schemes are also known for giving stable and sharp
discrete shock profiles. They have also a high order of accuracy,
typically second order or more, in smooth regions of the solution.

We perform the time update of system (\ref{hydro_system})
according to the following conservative algorithm:
\begin{eqnarray}
    {\bf u}_{i,j}^{n+1} = {\bf u}_{i,j}^{n}
    & - & \frac{\Delta t}{\Delta r}
    (\widehat{{\bf f}}_{i+1/2,j}-\widehat{{\bf f}}_{i-1/2,j}) 
\nonumber \\
& - & \frac{\Delta t}{\Delta \theta}
    (\widehat{{\bf g}}_{i,j+1/2}-\widehat{{\bf g}}_{i,j-1/2}) 
    + \Delta t {\bf s}_{i,j} \, .
\end{eqnarray}
\noindent
Index $n$ represents the time level and the time (space)
discretization interval is indicated by $\Delta t$ ($\Delta r, \Delta\theta$).  The
``hat" in the fluxes is used to denote the so-called numerical fluxes
which, in a HRSC scheme, are computed according to some generic
flux-formula, of the following functional form (suppressing index
$j$):
\begin{eqnarray}
  \widehat{{\bf f}}_{i\pm{1\over 2}} = \frac{1}{2}
          \left( {\bf f}({\bf u}_{i\pm{1\over 2}}^{L})  +
                 {\bf f}({\bf u}_{i\pm{1\over 2}}^{R}) -
          \sum_{\alpha = 1}^{5} \mid \widetilde{\lambda}_{\alpha}\mid
          \Delta \widetilde {\omega}_{\alpha}
          \widetilde {r}_{\alpha} \right) \, .
\end{eqnarray}
\noindent
Notice that the numerical flux is computed at cell interfaces
($i\pm1/2$).  Indices $L$ and $R$ indicate the left and right sides of a
given interface. Quantities $\lambda$, $\Delta\omega$ and $r$ denote the
eigenvalues, the jump of the characteristic variables and the
eigenvectors, respectively, computed at the cell interfaces according
to some suitable average of the state vector variables.  Generic
expressions for the characteristic speeds and eigenfields can be found
in \cite{font2}.  Our code has the ability of using different
approximate Riemann solvers: the Roe solver \cite{Roe81}, widely
employed in fluid dynamics simulations, with arithmetically averaged
states and the Marquina solver \cite{donat96}, which has been extended
to Relativity in \cite{donat98}. The computations presented here were
obtained using Marquina's scheme.

A technical remark: the equilibrium star is supplemented by a
low-density uniform atmosphere, which is necessary for computing
non-singular solutions of the hydrodynamic equations everywhere in the
computational domain. After each time-step we reset the atmosphere's
density and pressure to their initial values, avoiding unwanted
accretion of matter onto the star. The influence of the atmosphere is
thus restricted to the surface grid-cells.

\section{Pulsations of Non-rotating Stars}
\label{nonrot}

Since our code uses spherical polar coordinates, it can also be employed 
to study the evolution of non-rotating stars in 1-D. In the evolution 
of initially static non-rotating stars, we observe the following 
properties (note that our numerical grid is Eulerian):

\begin{enumerate}
\item Small-amplitude radial pulsations are triggered by the
  truncation errors of the finite-differencing scheme.
\item The radial pulsations are dominated by a set of discrete
  frequencies, which correspond to the normal modes of pulsation of
  the star.
\item The numerical viscosity of the finite-difference scheme damps
  the pulsations and the damping is stronger for the higher frequency
  modes.
\item The presence of a constant density atmosphere affects the 
  finite differencing at the surface grid-cells, which increases
  the numerical damping of pulsations and also causes a
  continuous but very small drifting of the density distribution.
\end{enumerate}

The initial amplitude of the radial pulsations and the small drift in 
density converge to zero at a second order rate with increasing resolution.
The value of the density in the atmosphere region has a large effect on the
damping of the pulsations. If it is too large, the damping is strong. 
To minimize this effect, we typically set the density of the
atmosphere equal to $10^{-6}$ times the density of the last grid point
inside the star.

\subsection{1-D evolutions}

\begin{figure}
\centering
\epsfig{file=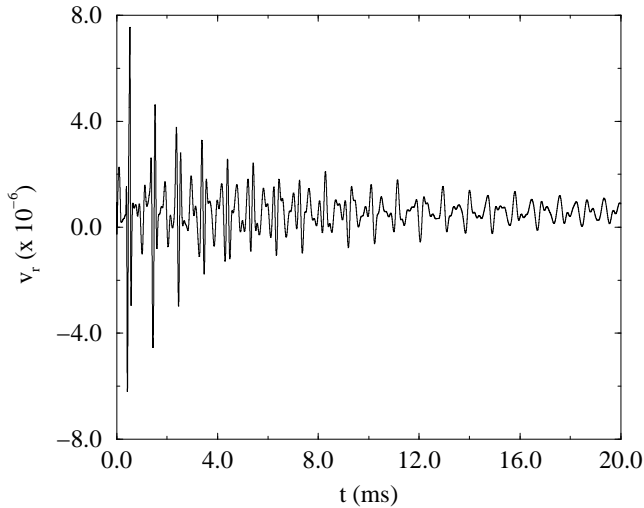,width=9cm}
\caption{Evolution of the radial velocity of an initially static, 
  non-rotating star, caused by the truncation errors of the
  finite-difference scheme. The radial pulsations are mainly a
  superposition of normal modes of the star.}\label{fig_nonrot1}
\end{figure}

We study the numerical evolution of a nonrotating $N=1.5$ relativistic
polytrope with $M/R=0.056$. The star is immediately set into radial
pulsation, triggered by the finite difference truncation errors. The
time-evolution of the radial velocity $v_r$, computed at 25\% of the 
star radius using a grid of 400 zones, is shown in Fig. \ref{fig_nonrot1}.
The vertical axis is dimensionless ($c=G=M_\odot=1$). The radial velocity
is initially a very complex function of time. As we will show next,
the pulsation consists mainly of a superposition of normal modes of
oscillation of the fluid.  The high frequency normal modes are damped
quickly and after 20ms the star pulsates mostly in lowest frequency
modes. Because these oscillations are caused only by the truncation
errors, the magnitude of the radial velocity is extremely small. As
shown in Fig.  \ref{fig_nonrot1} it is only a few times larger than
the non-zero residual velocity around which the star is oscillating.
This residual velocity converges to zero as second order with
increased resolution.

\begin{figure}
\centering
\epsfig{file=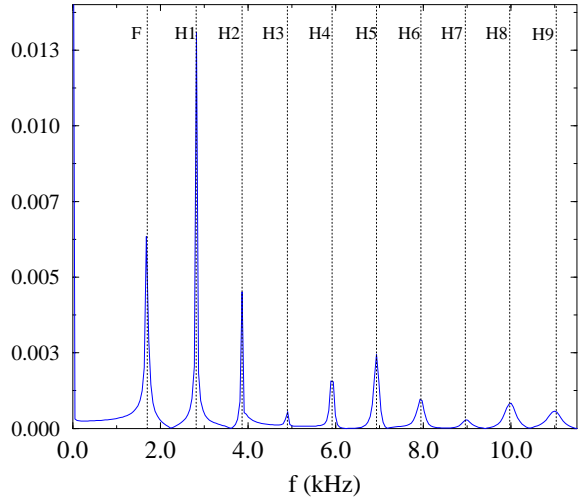,width=9cm}
\caption{Fourier transform of the evolution of the radial velocity in
  Fig. \ref{fig_nonrot1}. The frequencies are in excellent agreement
  with linear normal mode frequencies computed with an eigenvalue
  code. The units of the vertical axis are arbitrary.}
\label{fig_nonrot2}
\end{figure}

The small-amplitude radial pulsations in the non-linear, fixed
spacetime evolutions correspond to linear normal modes of pulsation in
the relativistic Cowling approximation, in which perturbations of the
spacetime are ignored. A Fourier transform of the density or radial
velocity time-evolution can be used to identify the normal mode frequencies.
Fig. \ref{fig_nonrot2} shows the Fourier transform of the radial velocity
evolution shown in Fig. \ref{fig_nonrot1}. The normal
mode frequencies stand out as sharp peaks on a continuous background.
The width of the peaks increases with frequency. The frequencies of 
radial pulsations identified from Fig. \ref{fig_nonrot2} are shown in
Table \ref{tab_radial1}.
   
To compare the obtained frequencies to linear normal mode frequencies,
we use a different code that solves the linearized relativistic
pulsation equations for the stellar fluid, in the Cowling
approximation \cite{McDermott,LindblomSplinter,YoshidaE}, as an
eigenvalue problem.  In Table \ref{tab_radial1} we present the results
of this comparison.  The typical agreement between frequencies
computed by the two methods is better than 0.5\% for the fundamental
$F$-mode and the lowest frequency harmonics $H_1-H_4$ and better than
0.8\% for the the higher harmonics $H_5-H_9$. This is a strong test
for the accuracy of the evolution code and our results can be used as
a testbed computation for other relativistic multi-dimensional
evolution codes.

\begin{table}[t]
\begin{center}
\begin{tabular}{*{4}{r}}
\multicolumn{4}{c}{}\\
\multicolumn{4}{c}{\large \bf Radial Pulsation Frequencies}\\ 
\multicolumn{4}{c}{}\\
\hline 
Mode & non-linear code (kHz) & Cowling (kHz) & difference \\[0.5ex]
\hline
\\[0.5ex]
$F$  &    1.703  &        1.697   &    0.3\%   \\[0.5ex]
$H_1$ &    2.820  &        2.807   &    0.5\%   \\[0.5ex]
$H_2$ &    3.862  &        3.868   &    0.02\%  \\[0.5ex] 
$H_3$ &    4.900  &        4.910   &    0.2\%   \\[0.5ex]
$H_4$ &    5.917  &        5.944   &    0.4\%   \\[0.5ex]
$H_5$ &    6.930  &        6.973   &    0.6\%   \\[0.5ex]
$H_6$ &    7.947  &        8.001   &    0.7\%   \\[0.5ex]
$H_7$ &    8.960  &        9.029   &    0.8\%   \\[0.5ex]
$H_8$ &    9.973  &        10.057  &    0.8\%   \\[0.5ex]
$H_9$ &   11.030  &        11.086  &    0.5\%   \\[0.5ex]
\end{tabular}
\vspace{3mm}
\caption{Comparison of small-amplitude radial pulsation frequencies
  obtained with the present non-linear evolution code to linear
  perturbation mode frequencies in the relativistic Cowling
  approximation. The equilibrium model is a nonrotating $N=1.5$
  relativistic polytrope with $M/R=0.056$.}
\label{tab_radial1}
\end{center}
\end{table}

\begin{table}[t]
\begin{center}
\begin{tabular}{*{4}{r}}
\multicolumn{4}{c}{}\\
\multicolumn{4}{c}{\large \bf Quadrupole Pulsation Frequencies}\\ 
\multicolumn{4}{c}{}\\
\hline 
Mode & non-linear code (kHz) & Cowling (kHz) & difference \\[0.5ex]
\hline
\\[0.5ex]
$f$   &    1.28   &  1.286  &    0.5\%         \\[0.5ex]
$p_1$  &    2.68   &  2.681  &    0.04\%         \\[0.5ex]
$p_2$  &    3.65   &  3.699  &    1.3\%         \\[0.5ex]
$p_3$  &    4.66   &  4.719  &    1.3\%         \\[0.5ex]
$p_4$  &    5.66   &  5.742  &    1.4\%            \\[0.5ex]
$p_5$  &    6.83   &  6.764  &    1.0\%       \\[0.5ex]
$p_6$  &    7.80   &  7.788  &    0.2\%       \\[0.5ex]
\end{tabular}
\vspace{3mm}
\caption{Comparison of small-amplitude quadrupole ($l=2$) pulsation 
  frequencies obtained with the present non-linear evolution code to
  linear perturbation mode frequencies in the relativistic Cowling
  approximation. The equilibrium model is a nonrotating $N=1.5$
  relativistic polytrope with $M/R=0.056$.}
\label{tab_quad1}
\end{center}
\end{table}

\subsection{2-D evolutions}

In a similar way, small-amplitude non-radial pulsations can be studied
with the present evolution code and the obtained frequencies can be
compared to perturbation results. We find that the truncation errors
of the finite difference scheme do not excite non-radial pulsations to
a sufficiently large amplitude compared to the amplitude of radial
pulsations, so that one cannot identify them accurately in a Fourier
transform. Instead, one has to perturb the initial configuration,
using an appropriate eigenfunction for each nonradial angular index
$l$. Such a perturbation can be constructed using the eigenfunctions
of linear pulsation modes, computed with the perturbation code in the
Cowling approximation. The frequencies of the non-radial modes are
then found from a Fourier transform of the time-evolution of the
velocity component $v_\theta$.

Table \ref{tab_quad1} shows a similar comparison as in Table
\ref{tab_radial1} for the quadrupole ($l=2$) pulsations of the same
$N=1.5$ relativistic polytrope. Since the non-radial modes have to be
computed on a 2-D grid, we cannot use resolutions as high as in the
1-D computations. For a small grid-size of $80\times80$ zones and a total
evolution time of 6ms, the agreement between frequencies computed by
the two methods is better than 1.4\% for the fundamental $f$-mode and
the $p$-modes $p_1-p_6$. For this grid-size, frequencies higher than
the $p_6$ mode could not be computed accurately, because the grid is
to coarse to resolve their eigenfunctions (higher harmonic
eigenfunctions have a larger number of nodes in the radial direction).

\section{Rotating Stars}

We now turn to the evolution of initially stationary, uniformly
rotating neutron stars. In these evolutions, we observe the same
qualitative properties as for non-rotating stars (section
\ref{nonrot}) and an additional important property: {\it the angular
  momentum of the star is not conserved at the surface layer}. This is
due to the fact that the velocity component $v_\phi$ of the fluid
has a maximum at the surface, while the numerical scheme (although
second-order accurate in smooth regions of the solution) is only
first-order accurate at local extrema. Moreover, the code evolves the
relativistic momenta, $S_i$, and the velocity components (as well as
the rest of ``primitive" variables) must be recovered through a root
finding procedure which involves dividing by the density.  At the
surface of the star (where the density is very small) this contributes
to obtaining less than second-order accuracy.

\begin{figure}
\centering
\epsfig{file=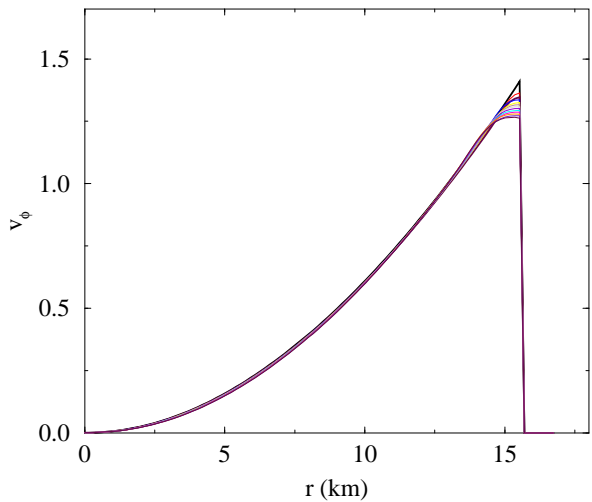,width=9cm}
\caption{Time-evolution of the velocity component $v_\phi$ of a rotating star 
  (see text for details). The accuracy is second order in the interior
  but only first order at the surface. This results in an angular
  momentum loss of the surface layers.}\label{f_rot1}
\end{figure}

A representative example of the evolution of a rotating star is
presented in Fig. \ref{f_rot1}, which shows the evolution, at
different times, of the velocity component $v_\phi$. The star is again a $N=1.5$
polytrope with the same central density as the non-rotating star
presented in section \ref{nonrot} and rotating at 74\% of the
mass-shedding limit at same central density.  The evolution was for
one rotation period on a $96\times60$ grid. The vertical axis is
dimensionless ($c=G=M_\odot=1$). The figure shows that the $\phi$-velocity
in the interior of the star remains close to its initial value, while
it decreases as a function of time in the outer layers. We find that
this is a generic property of the present numerical scheme for any
rotation rate and for any grid-size. By comparing evolutions with
different grid sizes, we verified that the loss of angular momentum at
the surface improves as first-order with resolution, while the
evolution of the $\phi$-velocity in the interior is second-order
accurate. However, as the evolution proceeds in time, the first-order
surface effect gradually affects the interior of the star.

\section{Quasi-radial Modes of Rotating Stars}

\begin{figure}
\centering
\epsfig{file=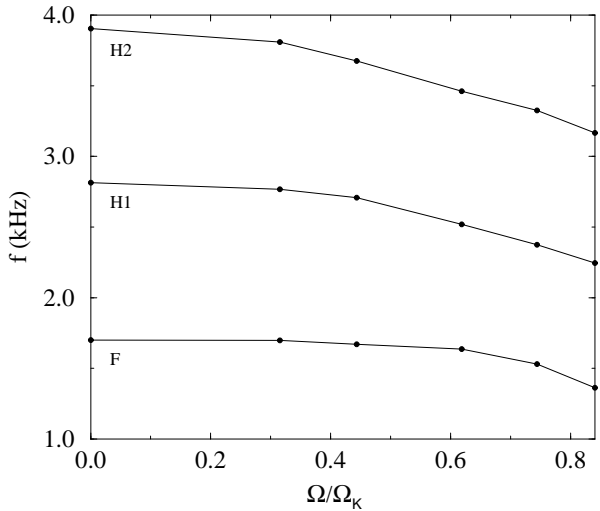,width=9cm}
\caption{Frequencies of the lowest three quasi-radial modes vs. 
  the ratio of angular velocity of the star $\Omega$ to the angular
  velocity at the mass-shedding limit $\Omega_K$, for the sequence of
  rotating relativistic stars in Table \ref{qr_tab}.}
\label{f_rot2}
\end{figure}

As a first application of our code, we compute quasi-radial modes
(i.e. modes that in the non-rotating limit reduce to radial modes) of
rapidly rotating relativistic stars in the Cowling approximation.
Previously, these modes have been computed for fully relativistic
stars only in the slow-rotation limit (but without the assumption of a
fixed spacetime) by Hartle \& Friedman \cite{Hartle} (see also
\cite{Datta}). We compute the three lowest-frequency quasi-radial
modes for a sequence of rotating stars of same central density. The
non-rotating member of the sequence is the non-rotating star of
section \ref{nonrot}. Table \ref{qr_tab} and Fig. \ref{f_rot2} show
our results for a low resolution grid of $100\times 80$ zones (note that our
computational grid assumes equatorial plane symmetry). For
this resolution we estimate the accuracy of the frequencies to be of
the order of $1-2$\%.

For the sequence of stars considered here, the frequencies of the
quasi-radial modes decrease with increasing rotation rate. This agrees
with previous slow-rotation computations which predict a decrease as
$\Omega^2$, where $\Omega$ is the angular velocity of the star. For fast
rotation, the change in the frequencies of quasi-radial modes is
affected by higher order terms in $\Omega$, because of the large
deformation of the equilibrium star. Also, for rapidly rotating stars
the quasi-radial mode frequencies are more ``closely packed'' than in
non-rotating stars.

\begin{table}[t]
\begin{center}
\begin{tabular}{*{4}{r}}
\multicolumn{4}{c}{}\\
\multicolumn{4}{c}{\large \bf Quasi-radial Pulsation Frequencies}\\ 
\multicolumn{4}{c}{}\\
\hline 
 $\Omega / \Omega_K$ &  F (kHz) & H1 (kHz) & H2 (kHz) \\[0.5ex]
\hline
\\[0.5ex]
0.0     &       1.71 &  2.82 &  3.90     \\[0.5ex]                      
0.32    &       1.70 &  2.77 &  3.81     \\[0.5ex]
0.44    &       1.67 &  2.71 &  3.68     \\[0.5ex]
0.62    &       1.64 &  2.52 &  3.46     \\[0.5ex]
0.74    &       1.53 &  2.38 &  3.33     \\[0.5ex]
0.84    &       1.36 &  2.25 &  3.17     \\[0.5ex]
\end{tabular}
\vspace{3mm}
\caption{Frequencies of quasi-radial modes for different values of the ratio of 
  angular velocity
  of the star $\Omega$ to the angular velocity at the mass-shedding limit
  $\Omega_K$, for a sequence of rotating relativistic stars of same
  central density. The non-rotating member of the sequence is the same
  as in Table \ref{tab_radial1}.}
\label{qr_tab}
\end{center}
\end{table}

\section{Discussion}

Our axisymmetric relativistic hydrodynamical code is capable to evolve 
rapidly rotating stars in a fixed spacetime. We find that, for non-rotating
stars, small amplitude oscillations have frequencies that agree with
linear normal mode frequencies in the Cowling approximation and we
compute the quasi-radial modes of rapidly rotating stars. 

Modern HRSC numerical schemes (as the ones used in our code), satisfying the
``total variation diminishing" (TVD) property \cite{H84}, are
second-order accurate in smooth regions of the flow, but only first-order
accurate at local extrema. In our rotating stars runs we find that this 
results in a loss of
angular momentum of the surface layers of the star, which gradually
also affects the interior of the star. This angular momentum loss only
vanishes as first-order with incresing resolution and we thus conclude
that for accurate long-term evolutions of rotating neutron stars it is
essential to use rather fine grids.  Furthermore, to reduce the
computational cost, one could use surface-adapted coordinates or
fixed-mesh refinement. It would also be interesting to see whether the
loss of angular momentum per rotation period will be significantly
smaller in a frame co-rotating with the star.  An alternative solution
to this problem, which we plan to investigate, could be the use of
``essentially non-oscillatory" (ENO) schemes, which maintain
high-order of accuracy even at local extrema \cite{HO87}.

All previous considerations are important for the study of the non-linear
dynamics of unstable toroidal oscillations ($r$-modes) in 3-D, which
have a long growth time and thus require highly accurate long-term
evolutions.

\section*{Acknowledgements}
We thank John L. Friedman, Curt Cutler, Philippos Papadopoulos and Tom
Goodale for helpful discussions. We also thank S. Yoshida for sending
us for comparison unpublished results on quasi-radial modes of
rotating stars in the Cowling approximation, computed with a linear
perturbation code.  J.A.F acknowledges financial support from a TMR
grant from the European Union (contract nr. ERBFMBICT971902). K.D.K.
is grateful to the Max-Planck-Institut f\"ur Gravitationsphysik
(Albert-Einstein-Institut), Potsdam, for generous hospitality.

\end{document}